\documentclass[aps,preprint,showpacs,showkeys]{revtex4}
\usepackage{amssymb,amsmath,epsf,graphicx}
\usepackage[ansinew]{inputenc}
\begin{document}
\title{Strong-field dipole resonance. I.  Limiting analytical cases}
\author{Christoph Uiberacker}
\email{christoph.uiberacker@unileoben.ac.at}
\affiliation{Institut für Physik, Montanuniversität Leoben, Franz-Josef Str.~18, 8700 Leoben, Austria}
 
\author{Werner Jakubetz}
\email{werner.jakubetz@univie.ac.at}
\affiliation{Fakult\"at f\"ur Chemie, Universit\"at Wien, W\"ahringer Str. 17, A-1090 Wien, Austria}
 
\begin{abstract}
We investigate population dynamics in $N$-level systems driven beyond the linear regime by a strong external 
field, which couples to the system through an operator with nonzero diagonal elements. As concrete example we 
consider the case of dipolar molecular systems. We identify limiting cases of the Hamiltonian leading to 
wavefunctions that can be written in terms of ordinary exponentials, and focus on the limits of slowly
and rapidly varying fields of arbitrary strength. For rapidly varying fields we prove for arbitrary $N$ that 
the population dynamics is independent of the sign of the projection of the field onto the dipole coupling. 
In the opposite limit of slowly varying fields the population of the target level is optimized by a 
{\it dipole resonance} condition. As a result population transfer is maximized for one sign of the field and 
suppressed for the other one, so that a switch based on flopping the field polarization can be devised. For significant sign dependence the resonance linewidth with respect to the field strength is small. In 
the intermediate regime of moderate field variation, the integral of lowest order in the coupling can be 
rewritten as a sum of terms resembling the two limiting cases, plus correction terms for $N>2$, so that a 
less pronounced sign-dependence still exists. 
\end{abstract}
%lev cross and opt pumping, other multiphoton, 
%strong fld excit opt transitions in Qantumsys: multiphoton 
\maketitle
 
\newpage 
 
\section{Introduction}
 
When strong few-cycle-, one-cycle- or sub-one cycle pulses \cite{you,krausz1,drescher,krausz2,persson}, or 
arbitrarily shaped pulses interact with atomic or molecular many-level systems, significant population 
transfer may occur within a fraction of an optical cycle \cite{paramonov,uib1,uib2,doslic2}. Concepts 
based on averaging over field oscillations \cite{keldysh}, frequencies and detuning lose their importance, 
while properties like the carrier-envelope phase (CEP) \cite{krausz1} and interactions with  
permanent or induced dipole moments of the system take precedence. 
 
We have previously investigated population transfer in a dipolar molecular system induced by one- and sub-one 
cycle pulses and found a strong dependence of the dynamics on the CEP and in particular on the sign of the 
projection of the electric field onto the difference of permanent dipole moments of the states involved in 
the reaction, in the following referred to as "sign-dependence" of the field \cite{uib1,uib2}. A similar 
sign-dependence for vibrational excitation by half-cycle pulses was found by Korolkov {\it et.~al.}
~\cite{paramonov}, but not related to permanent dipole moments. On the other hand, Do\v{s}li\'c et al.~
\cite{doslicmanz} and Naundorf {\it et.~al.}~\cite{naundorf} phenomenologically discuss a dipole-moment 
induced tunneling resonance in a 2-level system (2LS) under a dc field, but do not go into detail and do not 
address the case of general pulses. Note that earlier Thomas \cite{thomas86} had already given the 
analytical solution for the population dynamics of a 2LS under a constant field. Tati\'c and Do\v{s}li\'c 
\cite{doslic} describe an analogous tunneling process in a dipolar molecule driven by a long single field 
lobe, which they interpret as a distorted dc field. In a series of papers, Meath, Power, Brown and coworkers 
\cite{meathpower84,meathbrown1, meathbrown2,meathbrown3, brown1, brown2} consider the interaction of laser 
pulses with dipolar molecules within the rotating wave approximation \cite{rwa}, including also the case of 
a 2LS with a pulse and an added static field \cite{meathpower84}.

In the present paper we take up these points, generalizing the ideas of Tati\'c and Do\v{s}li\'c to distort a 
constant field to lobe-like pulses and putting them on a firm footing. Thomas in ref.~\cite{thomas86} did not 
address the dependence of the dynamics on the field strength and the resonance properties of the Rabi-type 
process. These can be related to the Stark effect \cite{messiah} inducing a change of the energy eigenvalues 
by coupling to the field, and to WKB-like arguments \cite{messiah,child} suggesting that "resonant" transfer 
between two levels should be maximized at conditions corresponding to degenerate eigenvalues. In this spirit 
we look for conditions for resonances and investigate the influence of the sign of the field; for the 
important point of constructing propagating pulses see below. 
 
In the domain of short strong pulses traditional tools, like the rotating wave approximation \cite{rwa} 
or Floquet theory \cite{floquet, hanggi}, become inapplicable, while methods like semiclassical strong field 
theory \cite{Dykhne1, Dykhne2, Davis, Krainov,Ivanov} are suitable approximations, although only so for large 
quantum numbers. Although exact representations of the population dynamics in strong fields have been 
discussed \cite{exactrep1,exactrep2}, due to their complexity it is hard to gain the physical insight 
required e.~g. for constructing a simple field that selectively populates the target level. 

%The results are in accord with the features of numerical simulations for few-level model systems described in a companion paper.
%In order to investigate the role of the CEP and the sign of the field we in first performed numerical 
%simulations of simple model $N$-level systems within reduced parameters and analysed dynamics for $N=2,3,4$ in 
%terms of detuning of levels and optimal population transfer \cite{uib_modelsys}. We found a strong field 
%resonance which is governed by a condition on the amplitude instead of the frequency as in perturbation theory. 
 
Rather than aiming at exact solutions we search for limiting cases that may lead to sufficiently general 
analytic results for dipole-moment driven dynamics in $N$-level systems (NLSs) in order to find criteria for 
effective population transfer. We first note that the reason why analytic solutions of the Schr\"odinger 
equation do not exist in the general case is rooted in the non-commutativity of the two operators in the 
Hamiltonian corresponding to the unperturbed energies and the coupling to the perturbation. This suggests that 
a way to obtain a solution in form of an ordinary exponential, in contrast to the usual time-ordered 
exponential \cite{messiah}, is to search for cases where in a suitable representation of the Hamiltonian one 
of the two matrices can be neglected, and copies of the remaining matrix at different times commute with each 
other. 
 
We do not further pursue the well-known cases of the weak field limit, which can be treated by perturbation 
theory \cite{messiah}, and the strong field limit \cite{Dykhne1, Dykhne2, Davis, Krainov,Ivanov}. Instead we 
concentrate on two alternative situations, which are characterised by slowly and rapidly varying fields of 
arbitrary strength. In relation to standard frequency-driven conditions both these limits would correspond to 
extreme detuning. The limit of rapidly varying fields can be qualitatively analysed for arbitrary $N$ by 
transforming to the interaction representation. The resulting integrals are elementary and we find that in 
this limit population transfer in NLSs does not depend on the sign of the field. However, for propagating 
pulses the possible population transfer is negligibly small.
 
The more interesting case is the one of slowly varying fields, where within certain time-intervals the field 
can be well approximated by a constant value. This case can be solved by the adiabatic approximation \cite{schwinger} 
and diagonalization. We show that in this limit a resonance emerges which enables effective population transfer 
and will be interesting for applications. This ''dipole-resonance'' determines the magnitude and sign of the 
field in contrast to the resonance condition on the frequency in usual spectroscopic transitions under weak 
fields. Furthermore, for 2LSs the reference case of a constant field has an exact 
analytical solution \cite{thomas86}, so that a combination of these results with the present allows a more 
comprehensive understanding of the dipole resonance.
%The most extreme situation in this limit is represented by a dc field with its associated Stark-energies. 
%Dipole resonance between two levels is achieved by applying a suitable field such that the diagonal parts of the Hamiltonian for these levels become equal~\cite{naundorf}.

%is capable of explaining the dynamics found in our previous work on the isomerization by one- and sub-one cycle pulses, 
%and which points to several interesting applications. 
 
In the intermediate case of moderate field variation no general analytic treatment is possible. In this case 
we use the contribution of lowest order in the interaction picture and show that it can be rewritten as a 
sum of two terms, each one representing one of the above limiting cases of field variation. Thus we obtain 
the result that a somewhat less pronounced dependence on the sign of the field may occur, which becomes 
manifest only after a certain ''induction period''. We extend our treatment of this case to 3-level systems 
(3LSs), whenever possible also addressing generalizations to the case $N>3$.

Propagating pulses show the property that in the far-field the time-average of its electric field goes to 
zero \cite{you,persson, bandrauk, mourou}. Such pulses are naturally obtained from vector potentials with 
the property $\lim_{t\to\infty} [A(t)-A(-t)] = 0$. More pragmatically, "effective half-cycle pulses" obeying 
the zero time-average can be designed as fields with one pronounced lobe balanced by long, but weak tails 
with opposite sign of the field strength or by series of small lobes \cite{persson}. When using single lobes, 
we anticipate that such small "side" lobes will make negligible contributions if their peak field strength is 
far from the resonance field strength.  This standard approach of working with single lobe fields is justified 
{\it a posteriori} by simulations on model systems \cite{uib_modelsys}. 
 
Our paper is organized in the following way. In section~\ref{theory} we present general developments of the 
theory, which lead to the specification of the limiting cases of "slowly varying" and "rapidly varying" fields. 
Section~\ref{Secslowvar} is devoted to the case of slow field variation, in which we include separate subsections on 
2LSs and NLSs. Next section~\ref{Secfastvar} investigates the theory for rapidly varying fields, and in the second 
subsection we combine the results for the limiting cases to obtain results for the intermediate case of field 
variation. In Sec.~\ref{summary} we give a summary of our investigations together with our conclusions. In three 
appendices we in turn discuss spectral properties of propagating pulses, estimate the magnitude of multiple integrals 
of the field, and give details of the derivation of the results for rapidly varying fields.

\section{Theory: General development}
\label{theory}
 
We consider a sequentially coupled, potentially branched NLS, representing {\it e.g.} vibrational levels of a molecule with a permanent dipole moment. Using the semiclassical dipole approximation, the Hamiltonian can be written in algebraic form, and the time-dependent Schr\"odinger equation becomes
\begin{equation} i\partial_t {c}_k = \left[\epsilon_k-\mu_{kk}E(t)\right] c_k-\sum_{l\ne k}\mu_{kl}E(t) c_l~. \label{schreq}
\end{equation}
The $c_k$, $k=1,\dots,N$, are the time dependent expansion coefficients of the eigenstates of the 
potential with eigenvalues $\epsilon_k$. $E(t)$ is the projection of the electric field onto the dipole operator with expectation values $\mu_{kl}$. The diagonal elements $\mu_{kk}$ represent the dipole moments.
 
To proceed we employ a Taylor expansion starting from the diagonal elements of the Hamiltonian, and switch to the interaction picture at $t=0$. Defining new coefficients
\begin{equation}
\phi_k(t):=\exp\left\{i\int_0^t{\rm d}t'\left[\epsilon_k-\mu_{kk}E(t')\right]\right\}c_k(t)~,
\end{equation}
we obtain
\begin{equation}
i\partial_t \phi_k(t)=\sum_{l}C_{kl}(t)\phi_l(t)~,
\end{equation}
where the time-dependent matrix ${\bf C}(t)$ is given by
\begin{eqnarray}
C_{kl}(t) & := & -\mu_{kl}E(t)\exp\left\{i\int_0^t{\rm d}t'\left[\Delta\epsilon_{kl}-\Delta\mu_{kl}E(t')\right]\right\}~, \nonumber \\
C_{kk}(t) & := & 0~.
\label{cmat}
\end{eqnarray}
Here we also used
\begin{equation}
	\Delta\epsilon_{kl}:=\epsilon_k-\epsilon_l \quad , \quad \Delta\mu_{kl}:=\mu_{kk}-\mu_{ll}~.
\end{equation}
Note $C_{kl}(t)$ is non-zero only for coupled pairs $k,l$. The solution to this problem is the time-ordered exponential \cite{messiah},
\begin{eqnarray}
\phi_k(t) & = & \sum_l\left[ \delta_{kl}-i\int_0^t{\rm d}t_1 C_{kl}(t_1)+
  (-i)^2\int_0^t{\rm d}t_1\sum_{l'}C_{kl'}(t_1)
\int_0^{t_1}{\rm d}t_2 C_{l'l}(t_2)+\dots\right]\phi_l(0) \nonumber \\
& =: & \sum_l (I^{(0)}_{kl}+I^{(1)}_{kl}+I^{(2)}_{kl}+\dots)\phi_l(0)~.
\label{texp}
\end{eqnarray}
In this notation $I^{(n)}_{kl}$ represents a term with an $n$-fold integral. 
 
In the following we consider a transition from an initial state i to a final state f, for which there exists a unique shortest coupled path with $s$ steps. From eq.~(\ref{texp}) we have
\begin{eqnarray}
\phi_{\rm f}(t) & = & \left[-i\int_0^t{\rm d}t_1 C_{\rm fi}(t_1)+
  (-i)^2\int_0^t{\rm d}t_1\sum_{l'}C_{{\rm f}l'}(t_1)
\int_0^{t_1}{\rm d}t_2 C_{l'{\rm i}}(t_2)+\dots\right]\phi_{\rm i}(0) \nonumber \\
& = & (I^{(1)}_{\rm fi}+I^{(2)}_{\rm fi}+\dots)\phi_{\rm i}(0)~,
\label{phi}
\end{eqnarray}
where $I^{(r)}_{\rm fi}$ is an $r$-fold integral corresponding to an $r$-step process, and hence $I^{(r)}_{\rm fi}=0$ for $r<s$. We define the population $P_k(t)$ of a given level $k$ by the square modulus of the wavefunction $\phi_k(t)$ for this level. Using $P_{\rm i}(0)=1$, the population of the target level f then becomes 
\begin{equation}
P_{\rm f}(t) := |\phi_{\rm f}(t)|^2 = \sum_{n\ge s} \left\{|I^{(n)}_{\rm fi}|^2 + 2\sum_{m>n} {\rm Re} \left[ 
(I^{(n)}_{\rm fi})^* I^{(m)}_{\rm fi} \right] \right\}~.
\label{pop}
\end{equation}
Terms of the form $(I^{(n)}_{\rm fi})^* I^{(m)}_{\rm fi}, m\ne n$, are interference contributions. 
 
The properties of the series of time-ordered integrals are well studied \cite{messiah}, yet so far no simplifications have been derived for the case of general perturbations $E(t)$, {\it e.g.} by removing the time ordering or by deriving expressions in terms of elementary functions. Note that not even $I^{(1)}_{\rm fi}$ can be treated analytically for general functions $E(t)$. Integration of the Schr\"odinger equation without going to the interaction picture is not any simpler, and again leads to time-ordered integrals. Note however, assuming the equation can be solved by diagonalization, that the interaction picture does not lead to the same eigenvalues than the Schrödinger picture. In order to obtain the correct dynamics, it is essential to diagonalize the Hamiltonian matrix of the original Schr\"odinger equation.

With this situation in mind we address the question which conditions on the parameters would lead to an analytic solution. Certainly in the absence of time-ordering eq.~(\ref{phi}) would become an exponential of the integral of the matrix $\mathbf{C}$, and therefore the level populations could be calculated explicitly. 
Time ordering arises from the noncommutability of the Hamiltonian matrices taken at different times. In order to obtain "simple results" we have to identify conditions under which these matrices {\it do} commute. This leaves us with the following cases:
\begin{enumerate}
	\item {\it Strong field limit}: the energies involved are much smaller than the diagonal contributions 
			from the field, $\epsilon_k << \mu_{kk}E(t)$. 
			The dynamics can be obtained by diagonalization of the dipole matrix. 
	\item {\it Weak field limit}: The off-diagonal terms in the Hamiltonian matrix are small. 
			This case can be treated by perturbation theory \cite{messiah}.
	\item {\it Slowly varying (adiabatic) field limit}: For every $t_0$ using the expansion of the field
			field $E(t)=\sum_{n=0}^{\infty} \partial_t^n E(t_0)(t-t_0)^n/n!$ we assume that 			
			$|E(t)-E(t_0)|<<|E(t)|$. In	this case the dynamics can be well approximated by a system 
			of piecewise constant, time-independent Hamiltonians.
	\item {\it Rapidly varying field limit}: Considering the propagation matrix in the interaction picture, 
			we find a formal analytic solution in case of rapidly varying fields.
\end{enumerate}
 
In the following we will concentrate on the last two cases. To this end, as a measure of the variation of the field with time we introduce $\Omega_{\rm min}$ and $\Omega_{\rm max}$ as a characteristic lowest and highest Fourier frequency of $E(t)$, suitably determined from the spectrum. Now we can distinguish two limiting cases, quantifying in turn the conditions in items 3 and 4 above, 
 
(a) (slowly varying) $\min_{\{k,l\}}|\Delta\epsilon_{kl}|>>|\Omega_{\rm max}|$,  
 
(b) (rapidly varying) $\max_{\{k,l\}}|\Delta\epsilon_{kl}|<<|\Omega_{\rm min}|$,\\
as we show in the following.
Here $\{k,l\}$ denotes all pairs of levels within the reaction path. In terms of frequency, (a) and (b) correspond to the two opposite regimes of large detuning.

\section{Slowly varying fields}
\label{Secslowvar}
 
\subsection{Implications for slowly varying fields}

In order to treat the limit of slow variation, we start by representing the field as an expansion around a fixed value $t_0$, 
\begin{equation}
E(t) = E(t_0)+\sum_{n=1}E^{(n)}(t_0)(t-t_0)^n/n! ~. 
\label{fieldexpansion}
\end{equation}
Due to the low frequency of the field we expect that 
\begin{equation}
\left| \sum_{n=1}E^{(n)}(t_0)(t-t_0)^n/n! \right| << |E(t_0)| 
\label{E_smallderiv}
\end{equation}
holds, i.e.~we deal with "almost constant" fields. In the following we show the implications of this relation on the parameters of the field.

We rewrite the general spectral representation, eq.~(\ref{spectresEgeneral}) from appendix \ref{App_spectpulse}, with respect to inversion symmetry, so that
\begin{equation}
E(t) = E(t_0) +  \frac{1}{\pi}\int_0^{\infty} {\rm d}\Omega S_u(\Omega)\sin[\Omega (t-t_0)] + S_g(\Omega)\{\cos[\Omega (t-t_0)]-1\} ~.
\end{equation}
Here $S_u$ ($S_g$) denotes the spectrum corresponding to functions of odd (even) symmetry, and we define the phase, introduced in appendix \ref{App_spectpulse}, as $\phi := -\Omega t_0$. From the Fourier representation we explicitly took out the constant term, given by 
\begin{equation}
E(t_0)=\frac{1}{\pi}\int_0^{\infty} {\rm d}\Omega S_g(\Omega) ~.
\label{E_constpart}
\end{equation}
 
Replacing the sine and cosine functions by their Taylor series leads to 
\begin{eqnarray}
E(t) & = & \frac{1}{\pi}\int_0^{\infty} {\rm d}\Omega \left\{ S_u(\Omega)[\Omega (t-t_0)-\Omega^3 (t-t_0)^3/3!+\dots] \right. \nonumber \\
	&& \left. + S_g(\Omega)[1-\Omega^2(t-t_0)^2/2!+\dots] \right\} ~, 
\label{EFourier}
\end{eqnarray}
and therefore from eq.~(\ref{fieldexpansion}) and (\ref{EFourier}), using $E^{(0)}(t_0):=E(t_0)$ for compact notation, we obtain the relation
\begin{equation}
|E^{(n)}(t_0)| = \frac{1}{\pi}\int_0^{\infty} {\rm d}\Omega S_R(\Omega)\Omega^n ~,
\label{Emoments}
\end{equation}  
in which $R$ denotes the type of representation the spectrum belongs to, $R=g$ for $n$ even and $R=u$ for $n$ odd. Due to the discussion in Appendix \ref{App_spectpulse} we can translate the notion of "almost constant" to the property that the spectrum of $E(t)$ is sharply peaked at a small value $\Omega_0$, even in case we include the switch term. The linewidth of this peak is given by $2\Omega_0/\alpha\pi\sqrt{\ln(2)}$, which goes to zero with $\Omega_0\to 0$. Assuming the spectrum is similar to a gaussian peaked at $\Omega=0$, from eq.~(\ref{Emoments}) we find   
$|E^{(n)}(t_0)| \approx (\Omega_0/\alpha\pi\sqrt{\ln(2)})^n |E(t_0)|$, which gives rise to the upper bound $|E^{(n)}(t_0)| \le \Omega_0^n |E(t_0)|$.  
Using this relation in eq.~(\ref{fieldexpansion}) we obtain a geometric series as a majorant for $|E(t)|$, which converges for $|\Omega_0(t-t_0)|<1$, so that we obtain
\begin{equation}
|E(t)| \le \frac{|E(t_0)|}{1-\Omega_0(t-t_0)} ~.
\label{slowvar}
\end{equation}

Investigating the dynamics within a given time-interval of length $T$, {\it e. g.} the length of the pulse, we deduce from eq.~(\ref{slowvar}) that the condition for slow variation, which is equivalent to  $E(t)\approx E(t_0)$, becomes
\begin{equation} 
T<<1/\Omega_0 ~.
\label{OmegaBound}
\end{equation}
For the approximation to be true this sets a limit on the integration time, so that to each slowly varying field there is a longest time interval beyond which higher derivatives of the field become significant.
 
Next we derive a necessary condition for eq.~(\ref{E_smallderiv}), which is often intuitively connected with slow variation. We obtain
\begin{eqnarray}
|E(t)-E(t_0)| & \approx & \left|E(t_0)\sum_{n=1}\Omega_0^n(t-t_0)^n/n!\right|
                   =\left|(t-t_0)E(t_0)\left[\Omega_0+\sum_{n=2}\Omega_0^{n}(t-t_0)^{n-1}/n!\right]\right| 		
                   \nonumber \\
				&\approx & |E'(t_0)(t-t_0)| ~,
\end{eqnarray}
where we used $|\Omega_0(t-t_0)|<<1$. From $|E'(t_0)(t-t_0)|\approx|E(t)-E(t_0)|<<|E(t_0)|$ it follows that the condition of slow variation is equivalent to
\begin{equation}
	T << \left|\frac{E(t_0)}{E'(t_0)}\right| ~,
\end{equation}   
which is necessary, but not sufficient for eq.~(\ref{E_smallderiv}). 

We pause for a moment to discuss a subtle point about our treatment. In general we are interested in results for situations where the field is zero before and after the pulse. The present limiting case is unable to deal with a small field strengths (at the tails of a pulse), because the derivatives will eventually exceed the field strength even for slow variation. However, the case can be extended to complete pulses if the tails do not contribute significantly to the population dynamics. In numerical studies \cite{uib_modelsys} we indeed find this to be the case for the 2LS, and for $N>2$ with some restrictions due to alternative transfer pathways.  Furthermore, by studying the adiabatic approximation of a general 2LS and non-adiabatic coupling (NAC) to first order we find that the slope of $E(t)$ in the tails of the pulse, which gives rise to the NAC, has hardly any effect during the rise of the field strength, but is responsible that during its decrease the population remains in the target level \cite{uib_adiab}.

\subsection{The two-level system}

We now address the simple case of a 2LS, where the concept of a "dipole resonance" is particularly clear, and where the reference analytical solution is available for a 2LS under a constant field as the ultimate limit of slow variation. This section is hence connected to the question if, and how, a resonant constant field may be deformed into a pulse so that maximum population transfer is maintained.

\subsubsection{Dipole resonance in the two-level system}
\label{s_dipres}

In the 2LS we have only one possible step from level 1 to 2 
(which in our notation now become i and f).
In order to use the results of this section for $N>2$ 
discussed below, we consider the integral $M_1$ corresponding to the step
i to f as a possible innermost integral of an arbitrary term in
eq.~(\ref{phi}) that is part of a transfer pathway in an $N$-level system. 

By using the approximation $E(t)\approx E(t_0)[1+(t-t_0)\Omega_0+{\cal
O}(\Omega_0^2)]$ we obtain
\begin{equation}
        \int_0^T{\rm d}t E(t) = E(T)\int_0^T{\rm d}t [1+(t-T)\Omega_0] = TE(T)[1
+ \frac{T\Omega_0}{2}]
                        \approx TE(T)
\end{equation}
for the integral in the exponent of $M_1$. This is similar to the
adiabatic approximation~\cite{messiah,schwinger}, where the field is
treated as constant when performing the diagonalization and integration of
the Hamiltonian, with subsequent substitution of $E(t)$ for $E$ in the
result. We get
\begin{eqnarray}
        M_1(t_1) & \approx & i\mu_{l_1 \rm i}E(t_1)\int_0^{t_1}{\rm d}t_{0}
\exp\left\{i\left[\Delta\epsilon_{l_1 \rm i}-\Delta\mu_{l_1 \rm
i}E(t_0)\right]t_0\right\} \nonumber\\
 & \approx & \frac{ \mu_{l_1 \rm i}E(t_1)
\left[\exp\left\{i\left[\Delta\epsilon_{l_1 \rm i}-\Delta\mu_{l_1 \rm
i}E(t_1)\right]t_1\right\} - 1 \right] }
                         { \Delta\epsilon_{l_1 \rm i}-\Delta\mu_{l_1 \rm i}E(t_1) }~.
                         \label{thedipoleresonance}
\end{eqnarray}
This equation clearly shows a resonance whenever $E(t)=\Delta\epsilon_{l_1
\rm i}/\Delta\mu_{l_1 \rm i}=:A_0$, which we term {\it dipole resonance}. We note that the resonance selects one sign of the field.
 
\subsubsection{The two-level system in a constant field}
\label{2LSconstE}

\begin{figure}
 \begin{center}
   		\includegraphics[scale=0.7]{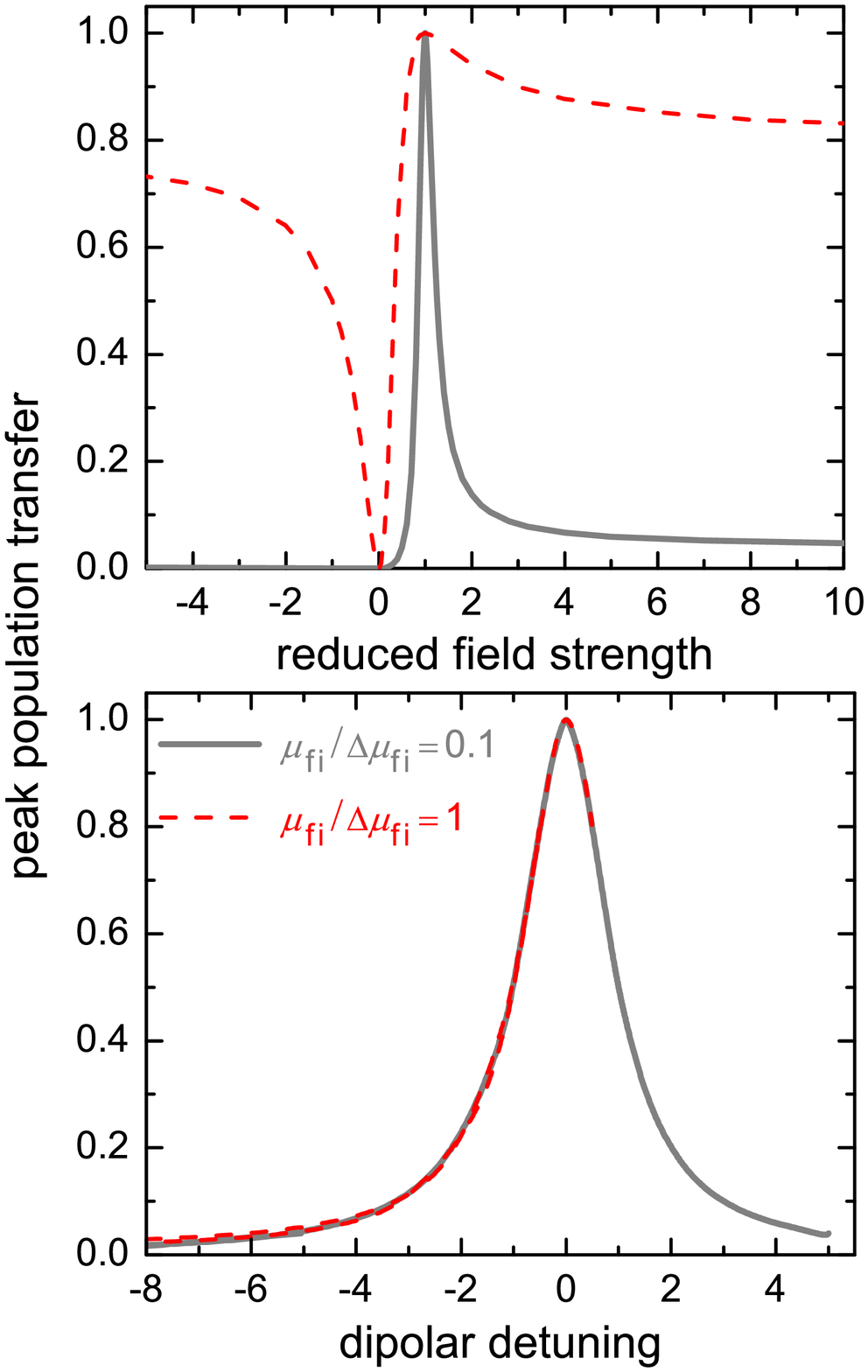}
   \caption{}
   \label{f_2lsPfmax}
%  Figure 1
 \end{center}
\end{figure}

The resonant behaviour seen in eq.~(\ref{thedipoleresonance}) is equally obtained for the 2LS under a constant field, as expected. As noted by Thomas \cite{thomas86}, the analytical solution for the evolution of population in the 2LS under a constant polarized field with (effective) field strength $E$ is governed by Rabi-dynamics. It is convenient to use reduced parameters, slightly different than the ones used in \cite{thomas86}, which are defined as
\begin{equation}
	d := \frac{\Delta\epsilon_{\rm fi}-\Delta\mu_{\rm fi}E}{2\mu_{\rm fi}E} \quad , \quad \theta := \frac{\mu_{\rm fi}E}{2}~t ~.
\end{equation}
We call $d$ the dipolar detuning, zero at dipole resonance, and $\theta$ is a generalized time. 
Using the initial condition $P_{\rm i} (t=0)=1$, we have
\begin{equation}
P_{\rm f}(\theta)= \frac{1}{1+d^2}\sin^2\left(\theta\sqrt{1+d^2}\right).
\label{rabidyn}
\end{equation}
We observe Rabi-like behaviour with $\theta$, where the dipole resonance is reflected by the prefactor containing the detuning in form of a Lorenzian. However, as shown in Fig.~1, variation of the "physical" tuning parameter $E$ does not result in a Lorenzian. We plot the maximal final population $P_{\rm f}^{max}(d)$, found by varying only $\theta$, against $E$ and $d$ in Fig.~\ref{f_2lsPfmax}. Note that $d=0$ represents the global maximum of $P_{\rm f}^{max}$ and in the limit of infinite field strength we only get $P_{\rm f}^{max}=\mu_{\rm fi}^2/(\mu_{\rm fi}^2+\Delta\mu_{\rm fi}^2/4)$, independent of the sign of the field.
Using the resonance field with inverted sign, $E=-\Delta\epsilon_{\rm fi}/\Delta\mu_{\rm fi}$, we obtain a maximal population in level f of $P_{\rm f}^{max}=\mu_{\rm fi}^2/(\mu_{\rm fi}^2+\Delta\mu_{\rm fi}^2)$. Immediately we conclude that significant sign dependence occurs only for $\mu_{\rm fi}<<\Delta\mu_{\rm fi}$.

Next we use this criterion to derive a condition of sign-dependence for slowly varying fields. We map the slowly varying field onto a constant field by averaging $E(t)$ and then combine this relation with the one for slow variation in eq.~(\ref{OmegaBound}). Assuming we constructed $E(t)$ such that the average field strength is on resonance we obtain from eq.~(\ref{rabidyn}) that the first instance of population inversion is given by the Rabi time $t=\pi/\mu_{\rm fi} A_0$. Inserting into eq.~(\ref{OmegaBound}) we conclude $\Omega << (\mu_{\rm fi}/\pi\Delta\mu_{\rm fi})\Delta\epsilon_{\rm fi}$. Hence we have two possibilities to fulfil this condition: either we make $\Omega$ very small or we search for systems with $\mu_{\rm fi}>>\Delta\mu_{\rm fi}$. In the latter case $\Omega$ need not be too small but there is no sign-dependence.
We note that only in the former case do we get sign-dependence for which we find $\Omega << \Delta\epsilon_{\rm fi}$ as a necessary condition, which relates system parameters to field parameters.
Note that, supported by fig.~\ref{f_2lsPfmax}, a significant sign dependence implies a significantly narrow resonance peak as a function of $E$.

%, which is identical to the one found in the intermediate case.
%We can relate this result to the limit derived in eq.~(\ref{OmegaBound}), and find in addition the condition for sign dependence in the intermediate case, see eq.~(\ref{nodetun}) and its discussion at the end of section~\ref{intermedtwolevel}.

\subsection{$N$-level systems}
 
In order to illustrate a general system with $s\ge2$, we use the example of a 
2-step process i$\rightarrow$b$\rightarrow$f from an initial state i via an
intermediate state b to the final state f in a 3-level system. 
 
We start by analyzing $I^{(2)}_{\rm fi}$, which is the simplest term that already reflects the added complexity of more than two levels.
$I^{(2)}_{\rm fi}$ is the term of smallest order in the coupling that contributes to the population of the target level. Explicitly we have
\begin{eqnarray}
I^{(2)}_{\rm fi} &=& (-i)^2\mu_{\rm fb}\mu_{\rm bi}\int_0^t{\rm d}t_1 \int_0^{t_1}{\rm d}t_2 E(t_1)E(t_2)  \nonumber \\
&&\times \exp\left\{i\left[\Delta\epsilon_{\rm fb}t_1+\Delta\epsilon_{\rm bi} t_2 
  - \int_0^{t_1}{\rm d}t'\Delta\mu_{\rm fb}E(t') 
- \int_0^{t_2}{\rm d}t'\Delta\mu_{\rm bi}E(t')\right]\right\}~. 
 \label{I2}
\end{eqnarray}
Consistent with the condition of slow variation of the field we approximate this integral in the following way, similar to
the procedure in Sec.~\ref{s_dipres},
\begin{eqnarray}
I^{(2)}_{\rm fi} &\approx& (-i)^2\mu_{\rm fb}\mu_{\rm bi}\int_0^t{\rm d}t_1 \int_0^{t_1}{\rm d}t_2  E(t_1)E(t_2) \nonumber \\
&&\times \exp\left\{i [\Delta\epsilon_{\rm fb}- \Delta\mu_{\rm fb}E(t_1)]t_1 + [\Delta\epsilon_{\rm bi}- \Delta\mu_{\rm bi}E(t_2)]t_2 \right\}~. 
 \label{I2appr}
\end{eqnarray}
From our discussion above we know that the tails of the pulse can be ignored and therefore we use $E(0)=0$ as the lower boundary of each integration. This gives
\begin{equation}
I^{(2)}_{\rm fi} \approx \mu_{\rm fb}\mu_{\rm bi}E(t)^2\frac{ \exp\left(i[\Delta\epsilon_{\rm fi} - \Delta\mu_{\rm fi}E(t)]t\right) }
{[\Delta\epsilon_{\rm fi} - \Delta\mu_{\rm fi}E(t)][\Delta\epsilon_{\rm bi} - \Delta\mu_{\rm bi}E(t)]}
\label{I2slow}
\end{equation}  
where $\Delta\epsilon_{\rm fi},\Delta\mu_{\rm fi}$ are differences between values at i and f, and 
$\Delta\epsilon_{\rm bi},\Delta\mu_{\rm bi}$ describe a transition from i to b. 
We assume for the moment that $E(t)\ne \Delta\epsilon_{\rm bi}/\Delta\mu_{\rm bi}$
holds, {\it i.e.} no resonance occurs for intermediate steps. Eq.~(\ref{I2slow}) then clearly shows a resonance 
at $E(t)=A_0=\Delta\epsilon_{\rm fi}/\Delta\mu_{\rm fi}$, in which case the population of the final state becomes 
\begin{equation}
P_{\rm fi}|_{E=A_0}(t) = \mu_{\rm fb}^2 \mu_{\rm bi}^2 A_0^4 t^2/[\Delta\epsilon_{\rm bi} - \Delta\mu_{\rm bi}A_0]^2 ~. 
\label{pop3LSres}
\end{equation}
We call this an {\it overall resonance} because it establishes a resonance between the initial and the final state across all intermediate steps.
 
We return to the case of simultaneous resonance of any number of intermediate steps (but not all of them). In our example this reduces to the single step i$\to$b, 
so that we obtain the condition $E(t)=\Delta\epsilon_{\rm bi}/\Delta\mu_{\rm bi}$.
Eq.~(\ref{I2appr}) then becomes
\begin{equation}
I^{(2)}_{\rm fi} \approx (-i)^2\mu_{\rm fb}\mu_{\rm bi}E(t)^2\int_0^t{\rm d}t_1 t_1\exp\left\{i\left[\Delta\epsilon_{\rm fb}t_1
- \int_0^{t_1}{\rm d}t'\Delta\mu_{\rm fb}E(t')\right]\right\}~.
\end{equation}
The remaining integral contributes significantly only if an additional  
resonance condition is fulfilled - in our example the resonance of step b$\to$f. 
Apart from the case that the field was shaped to attain the correct resonance condition of each step just 
at the time transfer occurs there, this is only possible for systems where all steps have the same 
$A_0$ ("dipole-harmonic" systems). 
For this case no distinction with respect to $N$ is necessary. In dipole-harmonic systems 
transfer to any "final" state will not be highly selective as {\it all} intermediate levels are on resonance.

Our analysis for a two-step process has a natural generalization to $s>2$. Dynamics via an overall resonance (direct coupling of i to f even if $\mu_{\rm fi}=0$) is direct and hence different from step-wise population transfer.  
From the results obtained above for the 2LS and the 3LS we conclude that population transfer 
is maximized when an "overall" dipole resonance condition is met. The resulting 
resonance field is determined by a ratio of the average of energy differences and the average 
of dipole moments along the shortest path $\textsl{P}$ between the initial and the final state,  
\begin{equation}
A_0=\frac{\sum_{i\in P} \Delta\epsilon_i}{\sum_{i\in \textsl{P}} \Delta\mu_i}=\frac{\Delta\epsilon_{\rm fi}}{\Delta\mu_{\rm fi}}~. \label{overall}
\end{equation}
We note the interesting detail that within this approximation the parameters of the 
intermediate levels have no influence on $A_0$ (which reflects overall resonance). However, $P_{\rm f}$ depends on the properties of the intermediate level by eq.~(\ref{pop3LSres}).

\section{Rapidly varying fields and the intermediate case}
\label{Secfastvar}

\subsection{Rapidly varying fields}
 
The limit of rapid field variation is defined by the condition $\max_{\{k,l\}}|\Delta\epsilon_{kl}|<< |\Omega_{\rm max}|$. We show in  Appendix \ref{Appfastvar} that this limit is applicable whenever the majority of the Fourier spectrum of $E(t)$ is located well beyond the largest energy difference from the initial level to any level along the transfer path. An example would be a single field lobe with sufficiently short duration (there is no immediate restriction on the number of lobes). The $n$ levels along the transfer path are denoted   
$\{l_0,l_1,\dots,l_n\}$ with $l_0={\rm i}$ and $l_n={\rm f}$. We investigate the first step from i to $l_1$ and begin with defining the auxiliary functions
\begin{eqnarray}
	a(x) & := & \exp(i\Delta\epsilon_{l_1\rm i}x) \quad , \quad  
		b(x):=E(x)\exp\left(-i\Delta\mu_{l_1\rm i}\int_0^{x}{\rm d}x'E(x')\right) \nonumber \\
	B(x) & := & \int_0^x {\rm d}x'b(x')
	\label{defsRapVar}
\end{eqnarray} 
Due to the oscillating kernel, repeated integrals of $b$ could be zero at isolated points. In order to keep our results general, we consider the repeated integral of order $k=n_0$ nonzero, while all integrals of lower order with $k<n_0$ are zero. Using $n_0$ partial integrations, in Appendix \ref{Appfastvar} we show the validity of the approximation 
\begin{eqnarray}
	M_1(t_1) &=& i\mu_{l_1\rm i}\int_0^{t_1}{\rm d}x a(x)b(x) \nonumber \\
	& = & (-1)^{n_0-1}i\mu_{l_1\rm i} \left(\left. a^{(n_0-1)}(x)I_B(n_0,x)\right|_0^{t_1} 
			-\int_0^{t_1}{\rm d}x a^{(n_0)}(x)I_B(n_0,x) \right) \nonumber \\
	& \approx &  i\mu_{l_1\rm i}(-\Delta\epsilon_{l_1\rm i})^{n_0-1} a(t_1)I_B(n_0,t_1) ~.
	\label{M1approxGeneral}
\end{eqnarray}
in the present limit. In eq.~(\ref{M1approxGeneral}) we use the definitions $I_B(k,t):=\int_0^t {\rm d}x I_B(k-1,x)$; $I_B(1,t)=B(t)$ as the $k$-th integral of $B(t)$, and $a^{(k)}$ denotes the $k$-th derivative of $a$. The properties of $k$-fold iterated integrals of the field, $I_k(E,t)$, are discussed in appendix~\ref{App_multintE}. 

Only two cases are relevant when considering pulses: the integral over the pulse may be zero (propagating pulse) or nonzero, leading to $n_0=2$ or $n_0=1$, respectively.

\subsubsection{$n_0=1$}
\label{Bnonzero}

Explicitly inserting $a$ and $b$ into $M_1$ and using eq.~(\ref{A_intbFac}), we obtain
\begin{eqnarray}
 M_1(t_1) \approx \frac{\mu_{l_1\rm i}}{\Delta\mu_{l_1\rm i}}\exp(i\Delta\epsilon_{l_1\rm i}t_1)
\left\{1- \exp\left[-i\Delta\mu_{l_1\rm i} \int_0^{t_1} {\rm d}t' E(t')\right]\right\}~.
\label{inta}
\end{eqnarray}
We use this expression in the kernel of the integral $I_{l_2\rm i}^{(2)}$ that represents the 2-step process (from i to $l_2$) and apply eq.~(\ref{M1approxGeneral}) once more to obtain
\begin{eqnarray}
 M_2(t_2) &\approx& \frac{\mu_{l_2 l_1}\mu_{l_1\rm i}}{\Delta\mu_{l_1\rm i}}\exp\left[i(\Delta\epsilon_{l_2 l_1}+\Delta\epsilon_{l_1\rm i})t_{2}\right]  \nonumber\\
       && \times   \left\{\frac{1}{\Delta\mu_{l_2 l_1}}\left[1-\exp[-i\Delta\mu_{l_2 l_1}\int_0^{t_{2}}{\rm d}t'E(t')]\right]  \right.  \nonumber\\
       && \left. - \frac{1}{\Delta\mu_{l_2 l_1}+\Delta\mu_{l_1\rm i}}\left[1- \exp[-i(\Delta\mu_{l_2 l_1}+\Delta\mu_{l_1\rm i}) \int_0^{t_{2}} {\rm d}t' E(t')]\right]\right\}~.
\label{intJnm1calc}
\end{eqnarray}
Proceeding in this way we note that $M_k$ contains a sum of terms with the same functional form as $M_1$, which however depend on $\Delta\epsilon_{l_j \rm i}$ and $\Delta\mu_{l_j \rm i}$ and correspond to a subpath of the reaction path from i to $l_j$, with $j\le k\le n$; for details see Appendix \ref{Appfastvar}.
 
The calculation of the target state population $P_{\rm f}$ involves taking the square modulus of a
sum of terms consisting of a real factor multiplied by a product of the complex quantities $p_k(x)$ and $f_{\beta\alpha}$, which are defined in eq.\ref{A_abbrev}. It follows that $P_{\rm f}$ is written as a sum of factors 
\begin{eqnarray}
\lefteqn{ (p_k f_{\alpha\beta})^* p_l f_{\kappa\nu} + p_k f_{\alpha\beta} (p_l f_{\kappa\nu})^* =  
 					2\left\{ \cos\left(\sum_{j=l}^{k-1} \Delta\epsilon_{l_{j+1} l_j}x \right)\right. }
 					\nonumber \\ 
&& - \cos\left[\sum_{j=l}^{k-1} \Delta\epsilon_{l_{j+1} l_j}x  - 
					w_{\beta\alpha}\int_0^{x}{\rm d}t'E(t')\right] - \cos\left[\sum_{j=l}^{k-1} 						\Delta\epsilon_{l_{j+1} l_j}x + w_{\kappa\nu}\int_0^{x}{\rm d}t'E(t')\right]  \nonumber\\
&& +  \left.\cos\left[\sum_{j=l}^{k-1} \Delta\epsilon_{l_{j+1} l_j}x - 			
				(w_{\beta\alpha}-w_{\kappa\nu})\int_0^{x}{\rm d}t'E(t')\right]\right\}  ~,
\label{popsummands}
\end{eqnarray}
multiplied by real numbers. Without loss of generality we used $n\ge k>l$.

Regarding eq.~(\ref{popsummands}), a few comments are in order.  
In case of a single initial state in a system with a unique path to the final state the sum over energy differences does not appear in eq.~(\ref{popsummands}), because direct contributions to the wavefunction and interference terms (path i to f augmented by loop paths from f back to f) all contain the same phase factor of energy differences. We arrive at the important conclusion that in this case the population of the final state does {\it not} depend on the sign of the external perturbation $E$. Note this result follows without any assumption about the number of field lobes. 
%However from eq.~(\ref{ineq_epst}) we see that it is essential that the maximum integration time (pulse duration) is determined by the smallest $\Delta\epsilon$ along the reaction path. 

If the initial population resides in more than one state, more than one path to the target level contributes to the dynamics, each one with its own energy difference and corresponding phase factor. In general, due to interference effects the dynamics will then depend on the sign of the field. This sign dependence however vanishes asymptotically in the strong field limit. 

\subsubsection{$n_0>1$}
\label{Bzero}

This case occurs whenever the field is supplied as a propagating pulse. It implies that the time-integral over the field  is zero \cite{mourou}, resulting in $S(0)=0$. This is a common situation in experiments which we therefore discuss here separately. 
In the general case we arrive at the estimate 
\begin{eqnarray}
 |M_1(t_1)| & = & {\cal O}\left(\frac{|\Delta\epsilon_{l_1\rm i}|^{n_0-1}}{\Omega_{min}^{n_0}}\right) ~.
\label{intM1general}
\end{eqnarray}
Equipped with this relation of general order we investigate population transfer at the end of a propagating pulse, $P_f(t\to\infty)$.
We can apply the approximation in eq.~(\ref{M1approxGeneral}) with $n_0=1$ for all integrations but the one corresponding to the last step, which contains $t\to\infty$ as an upper limit. The reason lies in the fact that $B(x)$ becomes zero at $x\to\infty$ (for all times after the pulse has passed). Hence we have to consider $n_0=2$ only in the integral representing the last step to level f. The same holds true for all possible interference terms. Thus from eq.~(\ref{A_nintEOrder}) we expect the maximum final population to be of the order $(\Delta\epsilon_{{\rm f}l_{s-1}}/\Omega_0)^2 << 1$ smaller than in case $n_0=1$. Due to the fact that this holds for arbitrary $s$, we conclude that a short, rapidly varying propagating pulse hardly transfers any population. It is important to recall that rapidly varying in our definition refers to the total field $E(t)$, independent of the envelope, and hence to a field of very high frequency. Like for $n_0=1$ we find that population transfer (regardless of its small magnitude) is independent of the sign of the field if we start from a single initial state.

\subsection{Intermediate case}
 
\subsubsection{2-level systems}
\label{intermedtwolevel}
 
If $|\Delta\epsilon_{\rm fi}|$ is comparable to both $\Omega_{\rm min}$ 
{\it and} $\Omega_{\rm max}$, we are in an "intermediate" regime concerning the two limits considered above. 
No analytic treatment is available, and we have to use other tools to analyse the integrals of the time-ordered series. 
We first note that due to the same functions $E(t)$ occuring in the exponent and in the 
factor multiplied with the exponential in the kernel of the integral, it is useful to use the identity
\begin{equation}
E(t) = -\left(g'_{\rm kl}(t)-\Delta\epsilon_{\rm kl}\right)/\Delta\mu_{\rm kl} \quad ,
\label{Esubst}
\end{equation}
with the time-derivative of the phase defined by
\begin{equation}
g'_{\rm kl}(t)=\Delta\epsilon_{\rm kl} -\Delta\mu_{\rm kl}E(t)~.
\label{gdef}
\end{equation} 
The indices $k,l$ are arbitrary and correspond to a given step. Using this relation in the integral representing the first step together with substitution leads to
\begin{eqnarray}
M_1(t_1) & = &  -\frac{\mu_{\rm fi}}{\Delta\mu_{\rm fi}}\left\{\exp\left\{i\left[\Delta\epsilon_{\rm fi} t_1-
\Delta\mu_{\rm fi}\int_0^{t_1}{\rm d}t' E(t')\right]\right\}-1\right . \nonumber \\
 & - & \left . i\Delta\epsilon_{\rm fi}\int_0^{t_1}{\rm d}t_0\exp\left\{i\left[\Delta\epsilon_{\rm fi} t_0
-\Delta\mu_{\rm fi}\int_0^{t_0} {\rm d}t' E(t')\right]\right\} \right\}~.
\label{nodetun}
\end{eqnarray}
The first term is equivalent to the expressions obtained in case (a). The second term can neither be calculated analytically nor be approximated as slowly varying in the whole domain of integration. However, from the discussion in section \ref{s_dipres} it is clear that for sufficiently short integration time we could apply the slowly varying approximation to the integrand. This leads in a natural way to the idea of partitioning the domain of integration into intervals. It only remains to find the intervals that show resonance. In the optimal case only one significant contribution remains and we can indeed replace the second term in eq.~(\ref{nodetun}) by its slowly varying approximation.

The integrand of the second term in eq.~(\ref{nodetun}) cannot be treated by the saddle point approximation (SPA) \cite{asymp} because $\hbar\to 0$ need not be true and in addition we integrate over a finite time-domain which results in contributions from the contour near $t_0=0$ and $t_0=t_1$. Elsewhere we will discuss a method of partitioning the domain of integration \cite{uib_seriesExp}. 

To find resonance points $\delta$, we search for a minimal first derivative of the modulus of the phase in the exponent. Around these points the variation of the phase factor in the integrand is slow and the resulting integral becomes large. Note this is a more general criterion than in the SPA, including all possible saddle points. 

In the following we abbreviate the phase difference between initial and final state with $g(t)$.
In order to use differentiation we note the bijective mapping of $|g'|$ to $g'^2$. 
This leads to 
\begin{equation}
\partial_t \left[{g'}(t)\right]^2|_{t=\delta} = 2{g'}(\delta)\partial_t {g'}(t)|_{t=\delta}=0 
~,
\label{lcond}
\end{equation}
and demanding a positive second derivative,
\begin{equation}
\partial_t^2 \left[{g'}(t)\right]^2|_{t=\delta} = 2\left[\partial_t {g'}(t)\right]^2|_{t=\delta}+
2{g'}(\delta)\partial_t^2 {g'}(t)|_{t=\delta}>0~,
\end{equation}
we obtain the centers $\delta$ of the resonance intervals with maximum population transfer. If $|{g'}(t)|>0$ for all times, then $\delta$ lies at an extremum of the field. Otherwise $g(t)$ becomes stationary, 
${g'}(\delta)=0$, so that $E(\delta)=\Delta\epsilon_{\rm fi}/\Delta\mu_{\rm fi}=:A_0$, 
which we denote true "dipole resonance". The latter case corresponds to the usual condition of the SPA. The above equations show that for oscillatory fields {\it population transfer can only be large for one given sign} within the period. The opposite sign corresponds to a maximum of $g'^2$ and yields hardly any transfer. In passing we note that the optimal case of a dipole resonance can only occur if $g(t)$ is not strictly monotonic. 
 
%When considering the Stark effect within classical electrodynamics, the field $A_0$ is leading to equal total energies of the initial and the final state, which relates to a dressed state picture. In a semiclassical view, e.g.~using path integrals, this can be interpreted as the adiabatic limit and is related to strong field theory \cite{Dykhne1, Dykhne2, Davis, Krainov,Ivanov,Vitanov}.
 
In general, we may obtain more than one solution $\delta$. As an example, for $E(t)=A\sin\omega t$ the solutions of eq.~(\ref{lcond}) that lead to maximum population transfer are given in Table \ref{tab1}.
\begin{table}
\begin{tabular}{c|c|c|c}
\hline
        & \phantom{m}$A_0>A>0$\phantom{m} & \phantom{m}$A_0>0>A>-A_0$\phantom{m} & \\
 \phantom{m}condition:\phantom{m}& or           & or          & $|A|\ge|A_0|$ \\   
        & $A_0<A<0$ & $A_0<0<A<-A_0$   &   \\  
\hline
solution: & $\delta_1=\frac{\pi}{2\omega}$ & $\delta_2=\frac{3\pi}{2\omega}$ & 
\phantom{m}$\delta_{3,4}=\frac{1}{\omega}\arcsin(\frac{A_0}{A})$\phantom{m} \\
\hline 
\end{tabular}
\caption{Possible solutions of eq.~(\ref{lcond}) in $[0,2\pi]$ corresponding to minima of 
$g'^2$ for $E(t)=A\sin\omega t$.}
\label{tab1}
\end{table}  
If $|{g'}(t)|$ remains sufficiently small between two solutions, the corresponding resonance intervals merge, and the dipole resonance condition is maintained for a particularly long time. If $E(t)$ consists of a single half-cycle lobe, its optimal amplitude $A$ should therefore be somewhat larger than $A_0$. 
%In a forthcoming article we discuss global approximation (regarding dynamics) of a given pulse by an optimal rectangle pulse, which gives a value for the ratio $A_0/A$~\cite{uib_globalApprox}.
 
From eq.~(\ref{nodetun}) it is apparent that the resonance condition has to be maintained 
for a time longer than $1/\Delta\epsilon_{\rm fi}$, or equivalently $1/t<\Delta\epsilon_{\rm fi}$, in 
order that the second term dominates the sign-independent first term.  
Combining this result with the condition of a slowly varying field, $\Omega t<<1$, (at least 
valid  around the interval of resonance) gives a condition relating properties of the field 
to the system parameters, namely $\min_{\{k,l\}}|\Delta\epsilon_{kl}| >> |\Omega_{\rm max}|$.
Note this relation is identical with the one presented in section \ref{2LSconstE}. Furthermore, it is opposite to the one for fast variation in section \ref{Secfastvar}. 
The sign dependence gradually disappears as the pulses 
become shorter. For very short pulses this case goes over to the limiting case (a).
 
\subsubsection{$N$-level systems}

For $N>2$, new aspects arise due to the fact that the single steps along the transfer path can have different resonance amplitudes. This necessitates the definition of what we call diagonal detuning. For a given (sub)path the diagonal detuning $d_p$ can be defined relative to a reference transfer path from initial to final state by comparing the phases,
\begin{equation}
	d_p(t) := g'_p(t) - g'_{\rm fi}(t) \quad .
\end{equation}
We realize that the diagonal detuning is time-dependent for general fields. At a true overall resonance we get $d_p(\delta)=g'_p(\delta)$ as the phase difference along the path $p$. Note in case of a dipole-harmonic system $d_p(\delta)=0$ for every subpath of the transfer path.

We again demonstrate the dynamics by discussing a two-step process. Similar to the treatment of the 
case $N=2$ we use eq.~(\ref{Esubst}) and substitution to split each of the two integrals in $M_2$ into a sum of 2 terms.
We first substitute for $E$ in the inner integral ,
\begin{eqnarray}
I^{(2)}_{\rm fi} &=& (-i)^2\frac{\mu_{\rm fb}\mu_{\rm bi}}{\Delta\mu_{\rm bi}}\int_0^t{\rm d}t_1  E(t_1) \exp\left[ig_{\rm fb}(t_1)\right] \nonumber \\
&& \times \left\{ \left( \exp\left[ig_{\rm bi}(t_1)\right]-1 \right) - i\Delta\epsilon_{\rm bi}\int_0^{t_1}{\rm d}t_2 \exp\left[ig_{\rm bi}(t_2)\right] \right\} \nonumber \\
& = & (-i)^2\frac{\mu_{\rm fb}\mu_{\rm bi}}{\Delta\mu_{\rm bi}} \left\{ \int_0^t{\rm d}t_1  E(t_1) \exp\left[ig_{\rm fi}(t_1)\right]\left( 1  
- \exp\left[-ig_{\rm bi}(t_1)\right] \right) - \right. \nonumber \\ 
&& - \left. i\Delta\epsilon_{\rm bi}\int_0^t{\rm d}t_1  E(t_1) \exp\left[ig_{\rm fb}(t_1)\right] \int_0^{t_1}{\rm d}t_2 \exp\left[ig_{\rm bi}(t_2)\right] \right\}~. 
\end{eqnarray}  
It is interesting to note that population transfer shows not only the expected resonance from the initial to 
the final level, indicated by $g_{\rm fi}$, but also a concurring one to the intermediate level. The latter 
resonance has the effect that in case of a harmonic system mainly the third term contributes and no resonance 
from the initial to the final level occurs. The third 
term is a correction term with no further possibility of simplification, which describes stepwise excitation.   
Using eq.~(\ref{Esubst}) with indices f and i for replacing $E(t_1)$ in the first two terms, and after writing the phases out explicitly we finally obtain 
\begin{eqnarray}
I^{(2)}_{\rm fi}(t) & = & \frac{\mu_{{\rm f}l_1}\mu_{l_1\rm i}}{\Delta\mu_{\rm fi}\Delta\mu_{l_1\rm i}}\left\{
\exp\left(i[\Delta\epsilon_{\rm fi} t-
\Delta\mu_{\rm fi}\int_0^{t}{\rm d}t' E(t')]\right)-1\right . \nonumber \\
 & - & i\Delta\epsilon_{\rm fi}\int_0^{t}{\rm d}t_{1}\exp\left(i[\Delta\epsilon_{\rm fi} t_{1}
-\Delta\mu_{\rm fi}\int_0^{t_{1}} {\rm d}t' E(t')]\right) \nonumber \\
 & + & i\Delta\mu_{\rm fi}\int_0^{t}{\rm d}t_{1}E(t_{1})\exp\left(i[\Delta\epsilon_{{\rm f}l_1} t_{1}
-\Delta\mu_{{\rm f}l_1}\int_0^{t_{1}} {\rm d}t' E(t')]\right) \nonumber \\
 & \times & \left. \left[1+i\Delta\epsilon_{l_1\rm i}\int_0^{t_{1}}{\rm d}t_{2}\exp\left(i[\Delta\epsilon_{l_1\rm i} t_{2} - \Delta\mu_{l_1\rm i}\int_0^{t_{2}} {\rm d}t' E(t')]\right)\right] \right\} ~.
\label{Nnodetun}
\end{eqnarray}

Apparently, an interpretation similar to the case $N=2$ above can be given. The first term does not
induce any dependence on the sign of the field. The second term gives rise to the overall resonance, with conclusions equivalent to $N=2$, if the appropriate differences of energies and moments are used.
The last term is a correction term and consists of two iterated integrals, each one with a resonance
field of the respective step. In case of a harmonic system this term will contribute along with the second term. In case of systems with small but nonzero diagonal detuning a complicated temporal behaviour with beatings is possible. In case of large diagonal detuning along the path the third term will be negligible and the dynamics should be similar to an effective 2LS.  
Generalizing to more than 2 steps, the number of correction terms will increase. 
In case of alternate paths only such correction terms can contribute to quantum interference effects. We conclude that in systems with large diagonal detuning fairly symmetric switching of population by an appropriate choice of the field strength is possible because no interference can occur. 
 
Together with the results obtained for the 2-level system in the intermediate regime we conclude that in systems with sufficient diagonal detunings the relation
\begin{equation}
A_0=\frac{\Delta\epsilon_{\rm fi}}{\Delta\mu_{\rm fi}}
\end{equation}
remains valid for fields with moderate variation. 
The quality of the approximation of independence of $A_0$ on the details of the path increases with increasing diagonal detuning. 
 
In all situations discussed up to now we have assumed that each $\Delta\mu$ relevant for $A_0$ is nonzero. 
Considering for the moment the case $\Delta\mu_{\rm fi}=0$, we see that $A_0$ will tend to
infinity. This expresses the fact that in this case a resonance cannot occur, and 
population transfer can only proceed through other mechanisms.
 
Elsewhere we will present numerical simulations of population transfer in 2LS and 3LS for pulses and constant field \cite{uib_modelsys}, which clearly show the qualitative features discussed above. 

\section{Summary and conclusions}
\label{summary}
 
Using the time-dependent Schrödinger equation, we analyse the population dynamics in $N$-level systems induced by strong and short field pulses. The operator of the system, which couples to the external field, is assumed to contain different nonzero diagonal elements. In the present article we investigate as an example a system with permanent dipole moments coupled semiclassiclally to an electric field via the dipole operator.  
 
We search for possible limiting cases of the Hamiltonian, for which the solution of the Schrödinger equation can be well approximated by ordinary iterated integrals without time-ordering. Using respectively the original Schrödinger picture and the interaction picture, we identify two new regimes relevant for strong fields, namely the two opposite limits of large frequency detuning.

In the limit of rapid variation of the field, compared to a characteristic energy difference of the system, we prove for arbitrary $N$ that the population dynamics is independent of the sign of $E(t)$. Furthermore, for propagating pulses no significant population can be transferred by the pulse. 

In the opposite limit of slow field variation, population dynamics is determined by a resonance originating from the diagonal matrix elements of the observable contained in the coupling. This resonance selects the amplitude and phase of the field. This is in sharp contrast to the well-known resonance in the perturbative regime that determines the frequency of the field as a difference of eigenvalues for a transition. As a result population transfer crucially depends on the sign of the projection of $E(t)$ onto the appropriate difference of permanent dipole moments whenever the resonance linewidth with respect to the field strength is sufficiently small. 

We also give a qualitative discussion of the intermediate case of moderate variation of the field, by analysing the contribution of lowest order to the population in the final level. For $N=2$ we rewrite the integral to a form consisting of a sign-independent and a sign-dependent term, each corresponding to one of the limiting cases discussed above. A significant sign dependence occurs only for field pulses with a duration longer than the inverse level spacing of the initial and final level. In case of $N>2$, added correction terms must be considered. 
The correction terms usually show no resonance but their number increases with increasing number of levels, weakening the effect of the resonance. 

The pronounced sign-dependence of the population dynamics for slowly varying fields would allow for ready control of branching between target states with different polarity and thus suggests an application of dipole-resonant population transfer in a setup acting as a molecular switch. % cite switches!! 
For few-cycle pulses, the behaviour of the intermediate case will also become manifest as a dependence of population transfer on the carrier envelope phase. 
%Our findings are in accord with the results from simulations on population transfer in dipolar model systems \cite{uib_modelsys} 
 
%As a result a strong dependence of the population dynamics on the carrier envelope phase emerges.
%Finally, with the present analysis we are able to fully understand the CEP effects seen in HCN$\rightarrow$HNC isomerization dynamics induced by few-cycle pulses \cite{uib1, uib2}. 
 
The nature of the coupling and the origin of the diagonal terms is not relevant to our analysis, and hence our findings should also hold in various problems of optical and magnetic population dynamics. Even the condition of existence of diagonal elements of the operator coupling to the field can be relaxed by considering induced quantities within an effective Hamiltonian of dressed states. This might explain phenomena like the asymetric escape of electrons as a non-linear field effect in atoms.
  
\begin{acknowledgments}   
We thank 
%Misha Ivanov for discussions on strong field theory and                 
J. N. L. Connor for valuable hints and comments on integrals of oscillatory functions.                                                               
This work was sponsored by the Austrian Science Fund within the framework of the Special
Research Program F016 "ADLIS". 
\end{acknowledgments} 

\appendix

\section{Spectral properties of propagating pulses}
\label{App_spectpulse}
 
An experimentally admissible propagating pulse is formed by shaping the vector potential $\mathbf{A}(t)$ under the restriction that the difference of $\mathbf{A}$ at the beginning and the end of the pulse is zero. We assume a gaussian vector potential and write $A(t)$ for the projection of the vector potential onto the current density to obtain
\begin{equation}
A(t) = -c \frac{A m(t)}{\Omega_0}\sin(\Omega_0 t + \phi) \quad , \quad m(t) = \exp\left[-\frac{ (t-t_{max})^2}{\sigma^2}\right] ~.
\end{equation}
Here $A$ denotes an amplitude and $\Omega_0$ the frequency. The half-width at half maximum
of $m(t)$ is given by $\sigma/\sqrt{\ln(2)}$. This keeps the discussion fairly general because an arbitrary smooth envelope could be well approximated by a finite sum of gaussians with appropriate parameters.
Using $E(t):=-\frac{1}{c}\partial_t A(t)$ as the corresponding projection of the field onto the dipole moment we obtain
\begin{equation}
E(t) = A m(t) \left[ \cos(\Omega_0 t + \phi) - \frac{2(t-t_{max})}{\Omega_0\sigma^2}\sin(\Omega_0 t + \phi) \right] ~.
\label{E_propGaussPulse}
\end{equation}
We adjusted $\sigma$ to accommodate $\alpha$ optical cycles in the full width at half maximum of the gaussian envelope, i.~e., $\sigma = \alpha\pi\sqrt{\ln 2}/\Omega_0$ (see also ref. \cite{uib1}).
 
Starting from the vector potential in form of a gaussian pulse, in eq.~(\ref{E_propGaussPulse}) we obtain two terms for the electrical field. The second term is usually denoted the {\it switch term}. We will call the first term the {\it principal term}.
We first derive the spectral representation for the principal term, which can be rewritten as
\begin{equation}
    E_P(t) = A\exp\left(-\frac{t^2}{\sigma^2}\right)\cos(\Omega_0 t + \phi) ~,
\end{equation}
by changing the origin of the time axis to $t_{max}$ and redefining $\phi$ at the same instance. Furthermore, we use $\Omega_0\ge 0$ without loss of generality. The spectrum can be calculated analytically from its definition,
\begin{eqnarray}
S_P(\Omega) & := & \int_{-\infty}^{\infty} {\rm d}t E_P(t)\exp(-i\Omega t) \\
                & = & A\sigma\frac{\sqrt{\pi}}{2} \left\{ \cos(\phi)\left[
                            \exp\left(-\frac{\sigma^2}{4}(\Omega-\Omega_0)^2\right) +
                            \exp\left(-\frac{\sigma^2}{4}(\Omega+\Omega_0)^2\right) \right] \right. \nonumber \\
                && \left. + i\sin(\phi)\left[ \exp\left(-\frac{\sigma^2}{4}(\Omega-\Omega_0)^2\right) -
                            \exp\left(-\frac{\sigma^2}{4}(\Omega+\Omega_0)^2\right) \right] \right\} \\
                & =: & A\sigma\frac{\sqrt{\pi}}{2} \left[ S_{P,1}(\Omega)\cos(\phi) + iS_{P,2}(\Omega)\sin(\phi)\right]~.
\label{spect_principal}
\end{eqnarray}
Clearly the spectrum consists of two parts, which are even ($S_{P,1}(\Omega)$) and odd ($S_{P,2}(\Omega)$) with respect to inversion symmetry.
 
In order to find the spectral representation for the switch term, we again change the origin of the time axis and redefine $\phi$ to obtain
\begin{equation}
    E_S(t) = -\frac{2 A t}{\Omega_0\sigma^2}\exp\left(-\frac{t^2}{\sigma^2}\right)\sin(\Omega_0 t + \phi)
\end{equation}
Note that the switch term contains a sine, but due to its prefactor belongs to the same representation with respect to inversion symmetry as the principal term. The spectrum can be calculated in a similar way as for the principal term,
\begin{eqnarray}
\lefteqn{ S_S(\Omega) := \int_{-\infty}^{\infty} {\rm d}t E_S(t)\exp(-i\Omega t)} \nonumber \\
                && = -A\frac{\sqrt{\pi}}{2\Omega_0} \left\{ \cos(\phi)\left[
                            (\Omega-\Omega_0)\exp\left(-\frac{\sigma^2}{4}(\Omega-\Omega_0)^2\right) -
                            (\Omega+\Omega_0)\exp\left(-\frac{\sigma^2}{4}(\Omega+\Omega_0)^2\right) \right] \right.        
                            \nonumber \\
                && \left. + i\sin(\phi)\left[    
                            (\Omega-\Omega_0)\exp\left(-\frac{\sigma^2}{4}(\Omega-\Omega_0)^2\right) +
                            (\Omega+\Omega_0)\exp\left(-\frac{\sigma^2}{4}(\Omega+\Omega_0)^2\right) \right] \right\} \nonumber\\
                && =:  -A\frac{\sqrt{\pi}}{2\Omega_0} \left[ S_{S,1}(\Omega)\cos(\phi) +    
                            iS_{S,2}(\Omega)\sin(\phi)\right]~.
\label{spect_switch}
\end{eqnarray}
With respect to inversion symmetry, the spectrum again contains an even part ($S_{S,1}(\Omega)$) and an odd one ($S_{S,2}(\Omega)$).
 
Using eq.~(\ref{spect_principal}) and (\ref{spect_switch}) we find the spectral representation of the field by inverse Fourier transformation,
\begin{eqnarray}
E(t) & := & \frac{1}{2\pi}\int_{-\infty}^{\infty} {\rm d}\Omega [S_P(\Omega)+S_S(\Omega)]
                        \exp(i\Omega t) \\
         & = & \frac{A}{2\sqrt{\pi}}\int_{0}^{\infty} {\rm d}\Omega \left\{        
                         \left[\sigma S_{P,1}(\Omega) - \frac{S_{S,1}(\Omega)}{\Omega_0}\right]\cos(\phi)\cos(\Omega t)
                         \right. \nonumber \\   
         &&            \left. - \left[\sigma S_{P,2}(\Omega) - \frac{S_{S,2}(\Omega)}{\Omega_0}\right]
                             \sin(\phi)\sin(\Omega t) \right\} \nonumber\\
         & \approx & \frac{A}{2\sqrt{\pi}}\int_{0}^{\infty} {\rm d}\Omega
                                     \left[ \sigma S_{P,av}(\Omega) - \frac{S_{S,av}(\Omega)}{\Omega_0} \right]
                                     \cos(\Omega t + \phi)~, \nonumber
\label{spectresE}
\end{eqnarray}
The approximation in the last line uses $S_{.,av}:=(S_{.,1}+S_{.,2})/2$ and rests on the fact that the even and odd spectral contributions from the principal and switch terms form peaks that have similar shape near $\Omega_0$.
Invoking the relation between frequency and energy, this approximation can be rewritten as
\begin{equation}
    E(t) \approx \frac{A\sigma}{2\sqrt{\pi}}\int_{0}^{\infty} {\rm d}\Omega
                             \left[ S_{P,av}(\Omega) - \frac{S_{S,av}(\Omega)}{\alpha\pi\sqrt{\ln 2}} \right]
                             \cos(\Omega t + \phi)  ~.
\label{spectresEapprox}
\end{equation}
 
The spectrum of the principal term shows a symmetric peak at $\Omega_0$, whereas the spectrum of the switch term changes sign at $\Omega_0$. It possesses a negative and a positive peak with positions
\begin{equation}
    \Omega_{S,peak} = \Omega_0 \pm \frac{\sqrt{2}}{\sigma} ~,
\end{equation}
obtained as the local extrema of $S_{S,av}(\Omega)$.
 
From the shape of the spectra and eq.~(\ref{spectresEapprox}) it becomes clear that the resulting pulse has its peak slightly below $\Omega_0$ and extends farther towards smaller values of $\Omega$ than larger ones, due to cancellation of the principal and switch term spectra. The cancelation introduces a cutoff at
\begin{equation}
 \Omega_c \approx \Omega_0 + \frac{\sqrt{2}}{\sigma} = \Omega_0\left( 1 +
                                         \frac{\sqrt{2}}{\alpha\pi\sqrt{\ln 2}}\right) ~.
\end{equation}
We estimate the position of the peak of the total spectrum and obtain the maximum at
\begin{eqnarray}
 \Omega_P & = & \Omega_0 \left( 1 + \frac{\sigma}{2} - \sqrt{ \frac{\sigma^2}{4} +
                             \frac{2}{(\sigma\Omega_0)^2}} \right) \\
                     & \approx & \Omega_0 \left( 1 - \frac{4\Omega_0}{\alpha^3\pi^3(\ln 2)^{3/2}} \right) ~.
\end{eqnarray}
The approximation is valid whenever $6\alpha^4/\Omega_0^2>>1$ holds and $\Omega_P$ is very close to $\Omega_0$ in this case. 
 
Apparently in eq.~(\ref{spectresE}) only a cosine turns up because in case $\phi=0$ the vector potential is an odd function of $t$ and the field will be of even symmetry. The most general case contains terms of both characters and can be represented by
\begin{equation}
    E(t) = \frac{1}{\pi}\int_{0}^{\infty} {\rm d}\Omega    
                \left[S_u(\Omega)\sin(\Omega t+\phi) + S_g(\Omega)\cos(\Omega t+\phi)\right] ~.
	\label{spectresEgeneral}
\end{equation}
In appendix \ref{App_multintE} we use this form to discuss multiple integrals of the field pulse. We note that we could remove the phase $\phi$ by taking $\phi=\Omega t_0$, obtaining the usual Fourier composition.
 
\section{Multiple integrals of oscillatory functions}
\label{App_multintE}
 
For the proof in section \ref{Secfastvar} we need to estimate the size of repeated integrals of the function $b(x):=i\mu_{l_1\rm i}E(x)\exp\left(-i\Delta\mu_{l_1\rm i}\int_0^{x}{\rm d}x' E(x')\right)$. Note in case of significant variation of $E(t)$ the phase factor is close to 1 as its exponent is small. Due to $|b(x)|\le|\mu_{l_1\rm i}||E(x)|$ an investigation of respective integrals of $E(x)$ can be used to find an upper bound.
 
We write the spectrum of $E(x)$ as a sum of an odd and even part, $S(\Omega):=S_u(\Omega)+S_g(\Omega)$, and consider the single integration problem, 
\begin{eqnarray}
\int_0^{t}{\rm d}x E(x) &=& \frac{1}{\pi}\int_{0}^{\infty} {\rm d}\Omega \int_0^{t}{\rm d}x 	
				\left[S_u(\Omega)\sin(\Omega x+\phi)+S_g(\Omega)\cos(\Omega x+\phi)\right]  \nonumber \\
&\le & \frac{1}{\pi}\int_{0}^{\infty} {\rm d}\Omega \frac{S(\Omega)}{\Omega}  ~.
\label{intEFourier}
\end{eqnarray}
Here $\phi$ denotes a possible phase and we used the well-known fact that the integral of  $\sin(\Omega x+\phi)$ cannot be larger than the contribution of half a period. However, already at this point it is clear that this estimate is applicable only if $\lim_{\Omega\to 0}S(\Omega)$ goes as $\Omega^n$ with $n\ge 1$. We will generalize and refine this estimate below.
 
We are interested in the repeated integral
\begin{eqnarray}
\lefteqn{ I_n(E,t) := \underbrace{\int_0^t\dots\int_0^{x_{n-1}}}_n {\rm d}x_1\dots{\rm d}x_n E(x_n) } \nonumber\\
  && = \frac{1}{\pi}\int_{0}^{\infty} {\rm d}\Omega \underbrace{\int_0^t\dots\int_0^{x_{n-1}}}_n 
  				{\rm d}x_1\dots{\rm d}x_n \left[S_u(\Omega)\sin(\Omega x_n+\phi)
  				+S_g(\Omega)\cos(\Omega x_n+\phi)\right] ~.
\label{nInt}
\end{eqnarray}
Using the following exact expressions,
\begin{eqnarray}
 \int_0^t {\rm d}x \sin(\Omega x+\phi) & = & -[\cos(\Omega t+\phi)-\cos(\phi)]/\Omega \nonumber\\
 \int_0^t {\rm d}x \cos(\Omega x+\phi) & = & [\sin(\Omega t+\phi)-\sin(\phi)]/\Omega ~,
\label{sincosInt}
\end{eqnarray}
we can perform the integrations over time. Iterating we obtain the "boundary term" from the lower limits of the $n$-fold integral, given by
\begin{eqnarray}
%B_1 & = & [S_u\cos(\phi)-S_g\sin(\phi)]/\Omega \nonumber\\
%B_2 & = & \{ S_u(\Omega)[\cos(\phi)\tau+\sin(\phi)] - S_g(\Omega)[\sin(\phi)\tau-\cos(\phi)] 	
%						\}/\Omega^2 \nonumber\\
%B_3 & = & \{ S_u(\Omega)[\cos(\phi)\tau^2/2+\sin(\phi)\tau-\cos(\phi)]  \nonumber \\	
% 		&&			 - S_g(\Omega)[\sin(\phi)\tau^2/2-\cos(\phi)\tau-\sin(\phi)] \}/\Omega^3 \nonumber\\
B_n & = & B_{n-1}\tau/\Omega + {\mathbf S}(\Omega)\cdot{\mathbf B}(n,\phi)/\Omega^n \\
{\mathbf S}(\Omega) & := & \left[ \begin{array}{cc} S_u(\Omega) & -S_g(\Omega) \end{array} \right] \\
{\mathbf B}(n,\phi) & := & (-1)^{int[(n-1)/2]}\left\{ \begin{array}{ll} 
											 \protect{[\sin(\phi) \quad -\cos(\phi)]^T} &  n ~{\rm even} \\ 
											 \protect{[\cos(\phi) \quad \sin(\phi)]^T} &  n ~{\rm odd} 
											\end{array}  \right.  ~,
\end{eqnarray}
when defining $\tau:=\Omega t$ and the spectral and boundary vectors ${\mathbf S}(\Omega)$ and ${\mathbf B}(n,\phi)$ respectively. 
%We observe that $I_n(E,t)$ will approach zero with $t\to 0$ as $t^n$. 
Using the boundary contributions we obtain the following general expression
\begin{eqnarray}
I_n(E,t) & = & 
					\frac{1}{\pi}\int_{0}^{\infty} {\rm d}\Omega S_u(\Omega)F_u(\Omega,t)-S_g(\Omega)F_g(\Omega,t) \\
F_u(n,\Omega,t) & := & (-1)^{int[(n+1)/2]}\left\{ \begin{array}{l} \sin_n(\Omega t+\phi)
										\quad n ~{\rm even} \\ 
									  \cos_n(\Omega t+\phi)\quad n ~{\rm odd} \end{array}\right. \nonumber \\
F_g(n,\Omega,t) & := & (-1)^{int[(n+2)/2]}\left\{ \begin{array}{l} \cos_n(\Omega t+\phi)\quad n ~{\rm even} \\ 
									 	\sin_n(\Omega t+\phi)\quad n ~{\rm odd} \end{array}\right. ~.
\label{nintEgeneral}	
\end{eqnarray}
where the odd and even functions $F_u$ and $F_g$ have been defined in terms of what we denote modified sine- and cosine-functions. The first few of these functions and the general expressions are
\begin{eqnarray} 
\sin_1(\tau +\phi) & = & \left[ \sin(\tau +\phi) - \sin(\phi) \right]/\Omega \nonumber\\
\cos_1(\tau +\phi) & = & \left[ \cos(\tau +\phi) - \cos(\phi) \right]/\Omega \nonumber\\
\sin_2(\tau +\phi) & = & \left[ \sin(\tau +\phi) - [\sin(\phi)+\cos(\phi)\tau ] \right]/\Omega^2 \nonumber\\
\cos_2(\tau +\phi) & = & \left[ \cos(\tau +\phi) - [\cos(\phi)-\sin(\phi)\tau ] \right]/\Omega^2 \nonumber\\
& \vdots & \nonumber\\
\sin_n(\tau +\phi) & = & \left[1 - \left. \sum_{j=0}^{n-1} \frac{\tau ^j}{j!}\frac{\partial^j}{\partial \tau ^j}\right|_{\tau =0}\right]\sin(\tau +\phi)/\Omega^n \nonumber\\ 
\cos_n(\tau +\phi) & = & \left[1 - \left. \sum_{j=0}^{n-1} \frac{\tau ^j}{j!}\frac{\partial^j}{\partial \tau ^j}\right|_{\tau =0}\right]\cos(\tau +\phi)/\Omega^n  ~.	
\label{modCosSin}
\end{eqnarray}
The modified function of order $n$ is constructed by removing the first $n$ terms of the power series expansion of the respective sine or cosine function.
 
In case $\tau>1$ we find an upper bound for these function as follows,
\begin{eqnarray} 
|\sin_n(\tau +\phi)| & = & \left| \left[\sin(\tau +\phi) - \sin(\phi)+\cos(\phi)\tau 	
														+\sin(\phi)\frac{(\tau )^2}{2} \dots \right]/\Omega^n \right| \nonumber\\ 
& \le & t^n\left| \frac{1}{(\tau )^n}+\sum_{j=0}^{n-1}\frac{1}{j!(\tau )^{n-j}} \right| ~.
\label{uboundBigTau}
\end{eqnarray}
Within this approximation, the same upper bound applies to $\cos_n(\tau +\phi)$, and for the absolute value of both modified functions of order $n$ we obtain the upper bound 
\begin{equation} 
( 1 + e )\frac{t^n}{\tau} ~,
%\left( 1+\frac{1-\tau^{-n-1}/(n+1)!}{1-\tau^{-1}} \right) 
%										= t^n\frac{ 2\tau^{n+1}-\tau^{n}-1/(n+1)! }{ \tau^{2n+1}-\tau^{2n} } ~.
\label{uboundBigTau2}
\end{equation}
by replacing each inverse power of $\tau$ by $1/\tau$ and using the monotonic behaviour of $\sum_j^n 1/j!$ with $n$. The limit $\tau\to\infty$ of the modified functions is smaller but shows the same power in $\tau$, 
\begin{equation} 
\lim_{\tau\to\infty}|\sin_n(\tau+\phi)| \le t^n{\cal O}([(n-1)!\tau]^{-1})	~.
\label{uboundBigTau3}
\end{equation}
%We note that especially in the case of small $n$, which is of central interest in \ref{Secfastvar}, the upper bound is the most accurate one.
 
For $\tau\to 0$ the right side of eq.~(\ref{uboundBigTau}) diverges like the $n$-th power whereas the modified functions stay finite, namely
\begin{eqnarray} 
\lim_{\Omega\to 0} \sin_n(\tau +\phi) & = & \frac{(-1)^{int(n/2)}t^n}{n!}\left\{ \begin{array}{l} 
											 \sin(\phi) + \frac{(-1)^{int[(n+1)/2]} }{n+1}\cos(\phi)\tau  + {\cal O}(\tau ^2) 
											 \quad n~{\rm even} \\ 
											 \cos(\phi) + \frac{(-1)^{int[(n+1)/2]} }{n+1}\sin(\phi)\tau  + {\cal O}(\tau ^2) 
											 \quad n ~{\rm odd} 
											\end{array}  \right.  \nonumber\\ 
\lim_{\Omega\to 0} \cos_n(\tau +\phi) & = & \frac{(-1)^{int[(n+1)/2]} t^n}{n!}
											\left\{ \begin{array}{l} 
											 \cos(\phi) + \frac{(-1)^{int[(n+2)/2]} }{n+1}\sin(\phi)\tau  + {\cal O}(\tau ^2)
											 \quad n ~{\rm even} \\ 
											 \sin(\phi) + \frac{(-1)^{int[(n+2)/2]} }{n+1}\cos(\phi)\tau  + {\cal O}(\tau ^2) 
											 \quad n ~{\rm odd} 
											\end{array}  \right. ~.	\nonumber\\
\label{modCosSinOmzero}
\end{eqnarray}
We note that terms of equal order of the series for the modified sine and cosine function differ only by a sign and replacing $\sin(\phi)$ with $\cos(\phi)$.
Therefore we define auxilary functions ${\cal S}_n(\tau),{\cal C}_n(\tau)$ to write
\begin{eqnarray} 
\sin_n(\tau +\phi) & = & \frac{(-1)^{int(n/2)}t^n}{n!}
											 \sin(\phi)\left( 1 - \frac{\tau^2}{(n+1)(n+2)} + \dots \right) \nonumber\\ 	
&&  + \frac{(-1)^{int[(n+1)/2]}t^n}{(n+1)!}\cos(\phi)\left( \tau 
											 - \frac{\tau^3}{(n+1)(n+2)(n+3)} + \dots \right) \nonumber\\
& =: & \frac{(-1)^{int(n/2)}t^n}{n!}\sin(\phi){\cal C}_n(\tau)+ 		
											\frac{(-1)^{int[(n+1)/2]}t^n}{(n+1)!}\cos(\phi){\cal S}_n(\tau) \quad .  
\label{modSinSum}
\end{eqnarray}
in case of even $n$, with similar expressions for odd $n$ and for $\cos_n(\tau +\phi)$.
From the properties of alternating series it is clear that for $\tau\le 1$ we have $|{\cal S}_n(\tau)|\le 1,|{\cal C}_n(\tau)|\le 1$. This leads again to a common upper bound 
\begin{equation}
	\frac{(n+2)t^n}{(n+1)!} \quad {\rm for}~\tau\le 1 ~.
	\label{uboundSmallTau}
\end{equation} 
for both modified functions.
 
We now collect the results from eq.~(\ref{uboundBigTau2}) and eq.~(\ref{uboundSmallTau}) to estimate the size of $I_n(E,t)$. We split the integral over $\Omega$ at $\tau = 1$ which yields
\begin{eqnarray}
\left| I_n(E,t) \right|  & \le & \frac{t^{n-1}}{\pi} \left | \frac{(n+2)}{(n+1)!}\int_{0}^{1} {\rm d}\tau 	
									S(\tau/t) + (1+e)\int_{1}^{\infty} {\rm d}\tau \frac{S(\tau/t)}{\tau} \right |	\\
	& < & 	\frac{t^n}{\pi}\left[\frac{n+2}{(n+1)!}+1+e\right]\left |\int_{0}^{\infty} {\rm d}\Omega 	
									S(\Omega) \right |	~.
\label{nintEEstimate}	
\end{eqnarray}
From the definition of the spectrum [see eq.~(\ref{spectresE})] and $E(t)$ of a pulse in eq.~(\ref{E_propGaussPulse}) we get $E(0)=A\cos(\phi)\le A$ and the estimate becomes
\begin{equation}
	\left| I_n(E,t) \right| < A t^n\left[\frac{n+2}{(n+1)!}+1+e\right]	~.	
	\label{nintEEstimate2}	
\end{equation}
 
\section{Rapidly varying fields}
\label{Appfastvar}
 
First we assume the approximation in eq.~(\ref{M1approxGeneral}) valid and work out the general structure of the terms in $\phi_{\rm f}$ by abbreviating the following quantities occurring in $M_n$,
\begin{eqnarray}
&& p_n(x) := \exp\left(i\sum_{j=0}^{n-1} \Delta\epsilon_{l_{j+1} l_j}x \right)\prod_{j=0}^{n-1}\mu_{l_{j+1} l_j} \nonumber\\
&& w_{\beta\alpha} := \sum_{j=\alpha}^{\beta-1} \Delta\mu_{l_{j+1} l_j} \nonumber\\
&& f_{\beta\alpha}(x) := 1-\exp\left[-iw_{\beta\alpha}\int_0^{x}{\rm d}t'E(t')\right] ~. 
\label{A_abbrev}
\end{eqnarray}
Now we can write the first few $M_n$ as
\begin{eqnarray}
M_1 &=& p_1\frac{f_{10}}{w_{10}} \nonumber\\
M_2 &=& p_2\left(\frac{f_{21}}{w_{21}w_{10}}-\frac{f_{20}}{w_{20}w_{10}}\right) \nonumber\\
M_3 &=& p_3\left[\frac{f_{32}}{w_{32}w_{10}}\left(\frac{1}{w_{21}}-\frac{1}{w_{20}}\right)-
\frac{f_{31}}{w_{31}w_{21}w_{10}} + \frac{f_{30}}{w_{30}w_{20}w_{10}}\right] ~. 
\label{A_Mseries}
\end{eqnarray}
It is apparent that there are exactly $n$ terms in each $M_n$. The increase of the number of terms with increasing subscript comes from the two parts contained in $f$, namely $1$ and the exponential, so that each term in $M_n$ can be interpreted as a transition from some level $l_j,j<n$ to $l_n$. We do not pursue the details of the calculation any further, as we are only interested in the qualitative behavior of $P_{\rm f}$. We just note that these factors can be organized in a scheme resembling Pascal's triangle.
 
Next we give the proof of the approximation in eq.~(\ref{M1approxGeneral}). We investigate the integral corresponding to the level $l_j$ for which the energy difference to the initial level is maximal. For convenience we define $\epsilon:=|\Delta\epsilon_{l_j\rm i}|=|a'(x)|$ ($a(x)$ is defined in eq.~(\ref{defsRapVar})) and $\mu:=|\Delta\mu_{l_j\rm i}|$.
 
In the present case we assume that the spectrum is peaked at a sufficiently high frequency $\Omega_0$, with the property $\Omega_0>>\epsilon$. In addition the width of the peak is assumed much smaller than $\Omega_0$.
In this case $\Omega\in[0,\epsilon]$ hardly makes any contribution and can be ignored for finding an upper bound to $| I_n(E,t) |$.
 
In the following we set the lower limit of integration to the moment $t_0$ when the field was switched on. Because the derivative of the field up to any order is zero at $t<t_0$ it is clear that the lower boundary does not make a contribution and $\sin_n(\Omega t+\phi)$ is replaced by $\sin(\Omega t+\phi)/\Omega^n$ and similar for the modified cosine. From eq.~(\ref{nintEgeneral}) we obtain the upper bound
\begin{equation}
	\left| I_n(E,t) \right| \le \frac{A}{\pi}\left| \int_{\epsilon}^{\infty} {\rm d}\Omega 
			\frac{S(\Omega)}{\Omega^n} \right | ~,	
\label{A_nintEBoundRapVar}	
\end{equation}
which is equal to the maximally attained value of the oscillating function $I_n(E,t)$. Noting that the peak of $S(\Omega)$ is much larger than its halfwidth we obtain the order estimate 
\begin{equation}
	\left| I_n(E,t) \right| =	\frac{A}{\pi}{\cal O}(\Omega_0^{-n}) ~,	
\label{A_nintEOrder}	
\end{equation}
which we use in the proof below.
 
Now we show that $\epsilon << \Omega_0$ indeed implies the approximation made in eq.~(\ref{M1approxGeneral}) for $n_0=1$ and extend to $n_0>1$ later. The approximation clearly is valid whenever
\begin{equation}
\left| \int_0^{t_1}{\rm d}x a'(x)B(x) \right| << \left| a(x)B(x) \right| ~
\label{A_slowEcond}
\end{equation}
holds; the abbreviations are defined in eq.~(\ref{defsRapVar}).
In order to motivate the strategy of the proof we note that multiple integrals of $b(x)$ are oscillating functions, that is, strictly the lower bound of their magnitude is zero. Therefore we cannot proceed to construct an upper/lower bound for the left/right side of eq.~(\ref{A_slowEcond}) in order to prove it. We then use the estimate eq.~(\ref{A_nintEOrder}), noting that the inequality eq.~(\ref{A_slowEcond}) becomes invalid in small intervals around isolated zeros of $B(x)$. However, these intervals occupy a fraction of ${\cal O}(\epsilon/\Omega_0)$ of the whole integration time and noting that within these intervals we have $n_0$ incremented by 1, which multiplies an extra factor $\epsilon/\Omega_0$ to the wavefunction, the contributions from these intervals to the population are by a factor ${\cal O}[(\epsilon/\Omega_0)^2]<<1$ smaller than from the remaining intervals. 
 
Using eq.~(\ref{A_nintEOrder}) we immediately get 
\begin{equation}
 \left| \int_0^{t}{\rm d}x a'(x)B(x) \right| = \frac{A}{\pi}{\cal O}\left(\frac{\epsilon}{\Omega_0^2}\right)  ~. 
\label{A_intBFac}
\end{equation}
 
Next we have to obtain a suitable lower bound to the right hand side of eq.~(\ref{A_slowEcond}). We note that introducing $z(x):=-\exp(-i\mu \int_0^{x} {\rm d}x'E(x'))$ we obtain 
${\rm d}z(x)/{\rm d}x= i\mu E(x)z(x) = b(x)$. $B(x)$ can then be integrated analytically to give
\begin{equation}
 B(x) = \frac{1}{\mu}
\left\{1- \exp\left[-i\mu \int_0^{x} {\rm d}t' E(t')\right]\right\}~.
\label{A_intbFac}
\end{equation}
The exponent evaluates to $\int_0^{x} {\rm d}t' E(t')={\cal O}(A/\pi\Omega_0)\approx {\cal O}(\epsilon/\Omega_0)<<1$ when we consider field strengths of the order of the one from the dipole resonance condition (see \ref{s_dipres}). Due to the fact that the length of $1-\exp(-i\phi)$ increases monotonically with increasing $\phi\in[0,\pi]$ we are safe to postulate
\begin{equation}
 \left|1- \exp\left[-i\mu\int_0^{x} {\rm d}t' E(t')\right]\right| \approx 
	\left|\mu\int_0^{x} {\rm d}t' E(t')\right| = \frac{A\mu}{\pi}{\cal O}\left(\Omega_0^{-1}\right)  ~.
\label{A_expfct_approx}
\end{equation}
Inserting this result in eq.~(\ref{A_intbFac}) we get
\begin{equation}
 \left|B(x)\right| = \frac{A}{\pi}{\cal O}\left(\Omega_0^{-1}\right)  ~,
\label{A_intbFac_approx}
\end{equation}
which tells us that $|I_B(1,t)|\approx|I_1(E,t)|$ is a good approximation in our case. 
 
Putting it all together we immediately arrive at the desired result,
\begin{equation}
	\epsilon << \Omega_0 ~. 
	\label{A_condn0eq1}
\end{equation}
 
In case of $n_0>1$ we note the relation $|I_B(k,t)|\le|I_k(E,t)|$ for any $k$, with decreasing difference $|I_k(E,t)|-|I_B(k,t)|$ for increasing $\Omega_0$. We again use the estimate from eq.~(\ref{A_nintEOrder}) for both sides of the inequality to arrive at
\begin{equation}	 
		\frac{\epsilon^{n_0}}{\Omega_0^{n_0+1}} << \frac{\epsilon^{n_0-1}}{\Omega_0^{n_0}} ~,
\end{equation}
which clearly is equivalent to eq.~(\ref{A_condn0eq1}) and therefore the approximation in eq.~(\ref{M1approxGeneral}) holds for general $n_0$.

%%%%%%%%%%%%%%%%%%%%%%%%%%%%%%%%%%%%%%%%%%%%%%%%%%%%%%%%70%%%%%%%

\newpage

{\bf Figure captions}

Figure 1. (Color online) Maximum population $P_{\rm f}^{\rm max}(d)$ in the final level of a 2LS under constant field, plotted against $E/A_0$ (upper panel) and against the dipolar detuning (lower panel). The levels are denoted i (initial) and f (final). The resonance field strength is set to $A_0=\Delta\epsilon_{\rm fi}/\Delta\mu_{\rm fi}=1$, and two values of the ratio $\mu_{\rm fi}/\Delta\mu_{\rm fi}$ are used as indicated. The maximum detuning is given by $\Delta\mu_{\rm fi}/2\mu_{\rm fi}$, leading to the (visible) right-hand cutoffs in the lower graph. 

\newpage  
 
%%%%%%%%%%%%%%%%%%%%%%%%%%%%%%%%%%%%%%%%%%%%%%%%%%%%%%%%70%%%%%%%

\end{document}